\documentclass{article} 
\usepackage{amssymb}
\usepackage{amsmath}
\usepackage{color}
\usepackage{graphicx}
\usepackage{verbatim}
\usepackage{nips12submit_e,times}
\usepackage{listings}
 
\definecolor{dkgreen}{rgb}{0,0.6,0}
\definecolor{gray}{rgb}{0.5,0.5,0.5}
\definecolor{mauve}{rgb}{0.58,0,0.82}
 
\title{Accelerating Inference: towards a full Language, Compiler and Hardware stack}

\author{
Shawn Hershey, Jeff Bernstein, Bill Bradley, Andrew Schweitzer, Noah Stein, Theo Weber, Ben Vigoda \thanks{Emails are first.last@analog.com.   All authors are at Lyric Labs at Analog Devices. }  }

\nipsfinalcopy 
\begin{document}
\maketitle
\begin{abstract}
We introduce Dimple, a fully open-source API for probabilistic modeling. Dimple allows the user to specify probabilistic models in the form of graphical models, Bayesian networks, or factor graphs, and performs inference (by automatically deriving an inference engine from a variety of algorithms) on the model.  Dimple also serves as a compiler for GP5, a hardware accelerator for inference. 
\end{abstract}

\section{Introduction}

To a large extent, probabilistic graphical models unify a great number of models from machine learning, statistical text processing, vision, bioinformatics, and many other fields concerned with the analysis and understanding of noisy, incomplete, or inconsistent data.

Graphical models alleviate the complexity inherent to large dimensional statistical models (the so-called curse of dimensionality) by dividing the problem into a series of logically (and statistically) independent components. By factoring the problem into subproblems with known and simple interdependencies, and by adopting a common language to describe each subproblem, one can considerably simplify the task of creating complex Bayesian models.  Modularity can be taken advantage of further by leveraging this modeling hierarchy over several levels (e.g. a submodel can also be decomposed into a family of sub-submodels).  Finally, by providing a framework which abstracts the key concepts underlying classes of models, graphical models allow the design of general algorithms which can be efficiently applied across completely different fields, and systematically derived from a model description.

These observations have driven people towards developing computer languages and APIs to naturally specify complex probabilistic models, and automatically derive a corresponding inference engine (with limited input from the user), drastically reducing the prototyping time for probabilistic modeling. Many such tools have appeared over the last few years; examples include Infer.NET [3],  the Bayes Net Toolbox [4], BUGS, and Church[2] \footnote{Stochastic Matlab and Church belong in fact to probabilistic programming, an even richer class of probabilistic models, where the model is described by an arbitrary computer program with calls to elementary random procedures.}.
\section{Dimple}

Dimple is an open source API which allows the user to construct models that can be written as factor graphs. The model specification is separated from the inference engine. Models can then be defined without a priori consideration of which algorithm to use (or even knowing how present algorithms work). The modular architecture makes the development of new inference engines and schedulers easy, and enables the developer of a new algorithm to immediately run and test it on the relevant examples and target benchmarks (Dimple provides a number of such benchmarks). Dimple fully supports undirected and directed factors, discrete and continuous variables, and has growing support for parameter estimation (including generalized Baum-Welch).  It supports infinite streaming chains but does not yet support the more general infinite graphs needed to support nonparametric Bayesian statistics, although some preliminary work has been done to integrate Stochastic MATLAB[7] (a variant of Church) and Dimple.  

A crucial feature of the Dimple architecture is that it is very easy for anyone to add new inference engines and apply them to any model expressible by Dimple.  It is also very easy to write a new model and choose to use any of the solvers (or increasingly, to mix and match solvers).
 
A second crucial feature of Dimple is that it serves as a front-end and compiler for hardware-accelerated inference. In parallel to developing Dimple, Analog Devices, Inc. is also a hardware company aiming to build general purpose accelerators for machine learning and probabilistic programming in particular. A result of those efforts is a special purpose chip called GP5 - an accelerator for sum-product and min-sum belief propagation in discrete factor graphs.

At the time of writing this document, Dimple and Infer.NET have similar features: an ability to express finite graphical models, multiple inference engines (including Gibbs and Sum Product), support for both discrete and continuous variables, and support for both directed and undirected graphs.  Unlike Infer.NET, Dimple does not yet provide an Expectation Propagation inference engine, but the architecture would support such an addition.  On the other hand, Dimple does provide advantages over Infer.NET, such as giving users the option to create their own inference engines and custom schedules (the order in which messages are passes or samples are generated).  Dimple supports other solvers not present in Infer.net : particle belief propagation, sample based belief propagation, among others.  Dimple and Church differ more deeply.  Church can be used to model a wider range of probabilistic models that include infinite probabilistic graphical models, but also more complex models.  Dimple, on the other hand, can provide more types of inference engines because of the very same restrictions on the class of models it can describe.  In some cases, Dimple handles undirected factors and hard constraints (such as parity checks) more naturally than Church does. 

\subsection{Code example}

If data is encoded with a Low Density Parity Check code and then corrupted, one can try to recover it using belief propagation.  We provide an example of MATLAB Dimple code for decoding.  A typical implementation from scratch might take a thousand lines of code [5].

\begin{lstlisting}
% Parity Check Matrix:
M = [1 1 1 0 0 0; 1 0 1 0 1 0;  0 0 1 1 0 1];
%Instantiate a Factor Graph
My_Factor_Graph=FactorGraph();
My_Factor_Graph.Solver = ÒSumProductÓ;
%Create random variables with a domain of [0 1]
My_Variables=Bit(size(M,2),1);
%The factor function
xorDelta= @(bits) mod(sum(bits),2) == 0;
%Iterate over check equations and add constraints from the check matrix.
for i=1:size(M,1)
	variable_indices = find(M(i,:));
	My_Factor_Graph.addFactor(xorDelta, My_Variables(variable_indices));
end
%Set the priors
My_Variables = a_vector_representing_observed_data;
%Run inference
My_Factor_Graph.solve()
%Display the posterior distribution 
disp(My_Variables.Belief)
\end{lstlisting}
\subsection{Selected features}

In this section we will highlight some of Dimple's features.  A more comprehensive set of features can be found in the documentation found on GitHub. [1]

\subsubsection{Modeler Features}

\begin{itemize}
\item Easy specification of undirected and directed potentials.  Options include:
\begin{itemize}
\item Dense factors: any arbitrary function of the variables connected to the factor.  The following code generates a simple graph with two variables, $a$ and $b$ and a factor function $f(a,b) = e^{(-|a-b])}$ where $a$ and $b \in \{1,2,...10\}$ and the prior on $a$ defines $P(a=10) = 1$.
\begin{lstlisting}
%Create Factor Graph
fg = FactorGraph();
%Define function to penalize distance
f = @(x,y) exp(-abs(x-y));
%create two variables
a = Discrete(1:10);
b = Discrete(1:10);
%Add an undirected factor encouraging sameness between the 
%variables.
fg.addFactor(f,a,b);
%Set a hard input for a (100% certainty of a 10)
a.Input = [zeros(1,9) 1];
%Run inference
fg.solve();
%Spit out the posterior belief of b
disp(b.Belief');

\end{lstlisting}

Here we see the resulting posterior probability of $b$ after running sum product.

\begin{lstlisting}
ans =

    0.0001
    0.0002
    0.0006
    0.0016
    0.0043
    0.0116
    0.0315
    0.0856
    0.2326
    0.6321
    
\end{lstlisting}
\item Sparse factors: specify only those factor function values which are nonzero.  Improves performance by orders of magnitude when some variables are deterministic functions of others, for example.  The following code defines a factor function $f(a,b) = (a=b)$, enforcing the constraint that $a=b$ with equal weight for every value.
\begin{lstlisting}
%Create Factor Graph
fg = FactorGraph();
%Create two variables
a = Discrete(1:4);
b = Discrete(1:4);
%Specify the domain indices of non-zero entries in the factor table.
indices = [0 0;
           1 1;
           2 2;
           3 3];
%specify the weights for each row in the index table
weights = [1 1 1 1];
%Create the factor
fg.addFactor(indices,weights,a,b);
%Set a hard input
a.Input = [zeros(1,3) 1];
%Run inference
fg.solve();
%Print out posterior
disp(b.Belief);
\end{lstlisting}
After running inference, the following result is produced.
\begin{lstlisting}
ans =

    0.0000    0.0000    0.0000    1.0000
\end{lstlisting}
\item Nested factor graphs: once created, a factor graph can be instantiated within larger factor graphs, increasing code modularity, modifiability, reusability, and readability, in a paradigm similar to object oriented programming.  The following code snippet creates a graph that defines a soft 4-bit xor.  The parent graph nests this graph twice to create a toy error correcting code with two constraints over 4 variables each.
\begin{lstlisting}
%%%%%%% Create the nested graph
b = Bit(4,1);
fourBitXor = FactorGraph(b);
c = Bit();
fourBitXor.addFactor(@xorDelta,[b(1:2); c]);
fourBitXor.addFactor(@xorDelta,[b(3:4); c]);
 
%%%%%%% Create the parent graph
fg = FactorGraph();
a = Bit(6,1);
fg.addFactor(fourBitXor,a([1 2 4 5]));
fg.addFactor(fourBitXor,a([2 3 4 6]));
 
%Set inputs and run inference
a.Input = ones(6,1)*.9;
fg.solve();
disp(a.Belief');
\end{lstlisting}

After running inference, we see that all bits are pulled more strongly towards 1.  Graphs can be nested recursively, although not infinitely.

\begin{lstlisting}
ans =

    0.9654    0.9886    0.9654    0.9886    0.9654    0.9654
\end{lstlisting}
\end{itemize}
\item Vectorized Graph Creation: Achieving fast speeds with MATLAB requires avoiding loops.  Dimple provides the ability to create large n-dimensional collections of variables as well as large numbers of factors without loops.  This feature is different than streaming line graphs described later in this paper.

The following examples show just a few ways to vectorize graph creation.
\begin{itemize}
\item Chains
\begin{lstlisting}
N = 100;
b = Bit(N,1);
fg = FactorGraph();
ft = FactorTable(rand(2),b.Domain,b.Domain);
ft.normalize(1);
f = fg.addFactorVectorized(ft,b(1:end-1),b(2:end));
f.DirectedTo = b(2:end);
\end{lstlisting}

This code snippet creates a directed Markov Model.  It will result in the following graphical model.  Circles represent variables and squares represent factors.  
\begin{figure}[h!]
\center
\includegraphics{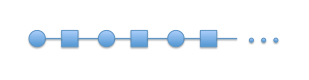}
\end{figure}
\item Grids
\begin{lstlisting}
N = 100;
D = 75;
b = Discrete(1:D,N,N);
fg = FactorGraph();
similarity = @(x,y) exp(-abs(x-y));
fg.addFactorVectorized(similarity,b(1:end-1,:),b(2:end,:));
fg.addFactorVectorized(similarity,b(:,1:end-1),b(:,2:end));
\end{lstlisting}
This code snippet creates the following graph and enforces similarity between adjacent variables.

\begin{figure}[h!]
\center
\includegraphics[scale=0.8]{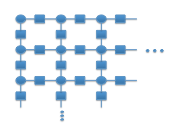}
\end{figure}

\item Fully Connected Layers: There are many ways to use the addFactorVectorized feature.  The Dimple documentation provides more detail.  However, we'll include just one more example to illustrate the flexibility of this feature.  The following code generates a graph with two layers where each node in layer 1 is connected to every node in layer 2.

\begin{lstlisting}
N = 100;
layers = Bit(N,2);
fg = FactorGraph();
similarity = @(x,y) exp(-abs(x-y));
fg.addFactorVectorized(similarity,repmat(layers(:,1),1,N),repmat(layers(:,2)',N,1));
\end{lstlisting}

The following image represents the graph.  Factors are omitted for readability.
\begin{figure}[h!]
\center
\includegraphics[scale=0.85]{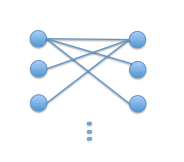}
\end{figure}
\end{itemize}
\item Streaming line graphs: many graphical models have a 1d structure (for instance, dynamic bayesian networks, kth order Markov Chains, HMMs and all varieties of Kalman filters).  Dimple provides the ability to create a template that is repeated over a specified window size.  Data can then be streamed through this window. For predictive inference, this allows effectively to use very large graphs while only using the memory required to carry sufficient statistics (i.e.\ the memory usage does not grow with the size of the line graph).  

The following code creates a Markov Chain.  The $rf.BufferSize=2$ code snippet specifies that the sliding window should hold two instances of the nested graph in memory so that some smoothing is performed in addition to prediction.

\begin{lstlisting}
b = Bit(2,1);
ng = FactorGraph(b);
similarity = @(x,y) exp(-abs(x-y));
ng.addFactor(similarity,b(1),b(2));

fg = FactorGraph();
b = DiscreteStream(0:1);
rf = fg.addRepeatedFactor(ng,b.getSlice(1),b.getSlice(2));
rf.BufferSize = 2;
N = 100;
dds = DoubleArrayDataSource(repmat([1 0],N,1));
b.DataSource = dds;

fg.initialize();
while true
    
    %The "false" argument tells Dimple not to initialize the 
    %messages so that as the rolled up graph progresses 
    %we don't lose information from the past.
    fg.solve(false);
    
    %Display belief of current variable
    b.FirstVar.Belief
        
    if ~fg.hasNext()
        break;
    end
    fg.advance();
end
\end{lstlisting}

\item Cluster belief propagation - loopy belief propagation can be improved by considering joint distribution over groups of variables (instead of marginal distributions); the graph can also be made less loopy by converting two or more factors into a single, larger factor. 
\item Multivariate Gaussians - Dimple provides support for Belief Propagation with Multivariate Gaussians with linear and nonlinear operator factors.  The Dimple documentation provides more detail.  Dimple also contains a set of demos, which include a Kalman filter.
\end{itemize}

\subsubsection{Architecture and Inference Features}

\begin{itemize}
\item Speed - Dimple's Belief Propagation engine is well optimized for many problems and we intend to optimize it further for others.
\item Open source - Lyric has made Dimple open source with the hope of contributing to the probabilistic programming community; how the community uses Dimple will also help guide the development of future features, improvements, and hardware.
\item MATLAB and Java API, extendable to any language on top of JVM - Dimple is written mostly in Java.  This was chosen as a trade-off in speed vs.\ ease of development.  A MATLAB front end is provided for convenience.
\item Customizable - The modular architecture allows all inference behavior to be customizable; one can even write entirely new inference engines.
Existing inference engines include Sum Product, Min Sum, a k-best approximation for Sum Product and Min Sum, Gaussian BP, Particle BP, and Gibbs sampling. As an example of customization, Dimple provides specialized factors for efficiently performing inference over variables with a finite field domain.
\end{itemize}
\section{Hardware acceleration}

\subsection{GP5}
As of this writing, Lyric Labs' first prototype of an inference accelerator, GP5, will be available in the next few months.  All acceleration estimates are based on the GP5 simulator, which is complete but not yet released publicly.  GP5 accelerates discrete Sum Product, Min Sum, and more generally, performs highly efficient weighted sparse tensor inner products.

The GP5 is optimized to accelerate inference on graphs with large factor tables.  A factor function is defined over the cartesian product of the domains of the connected variables; as a result, complexity of the factor inference is of order $O(nd^n)$ where d is the domain size of the connected variables and n is the degree of the factor. GP5 is aimed at optimizing inference in factor graphs where that factor complexity is relatively high. 

The actual execution time on the GP5 is determined by a collection of complicated and interlocking factors: I/O between the host processor and the GP5, various memory sizes on the GP5 relative to the message and factor table sizes of the factor graph, the scheduling of messages on the factor graph, the degree of each factor, and the various asynchronous sub-processors on the GP5.  Their interactions may produce computational bottlenecks that prevent the user from achieving peak performance.

The GP5 supports factor nodes that can connect to up to 16 variables and variables can connect to up to 256 factors.  The factor table cache is 256KB and the maximum domain size of variables is 4096.   Data can be moved in and out of the GP5 at 18Gb/s, the chip will be $36mm^2$, and draws between 1 and 2 watts of power.

In a binary image denoising benchmark[6] (variables with binary domains, factors with degree 16 connecting 4 by 4 image patches), GP5 is estimated to be 3200x faster than Dimple running on an intel i7, for a fraction of the power. The speedup is achieved by a large number of architectural optimizations, as well as the utilization of multiple small cores to parallelize inference on a single factor. Future versions could tile those cores to further enhance parallelism (including running several factors in parallel).  This is the best acceleration we have achieved in any of our benchmarks.

In the case of a stereo vision benchmark[6], the GP5 is estimated to be 100x faster than Dimple running on an intel i7.  The sterevision acceleration is worse than the denoising benchmark acceleration because of the small degree of the factor nodes in the stereo vision demo.

It is challenging to describe an expected acceleration across the range of models that can be used with the GP5 because of the complexity of the interlocking factors.  As a general guide, the larger the factor table and the larger the degree, the more acceleration the GP5 can achieve.  The GP5 simulator, however, can provide a fairly accurate estimation of the expected acceleration.

The following image shows the output of the Instruction Level Profiling Tool included with the GP5 simulator.

\includegraphics[scale=0.4]{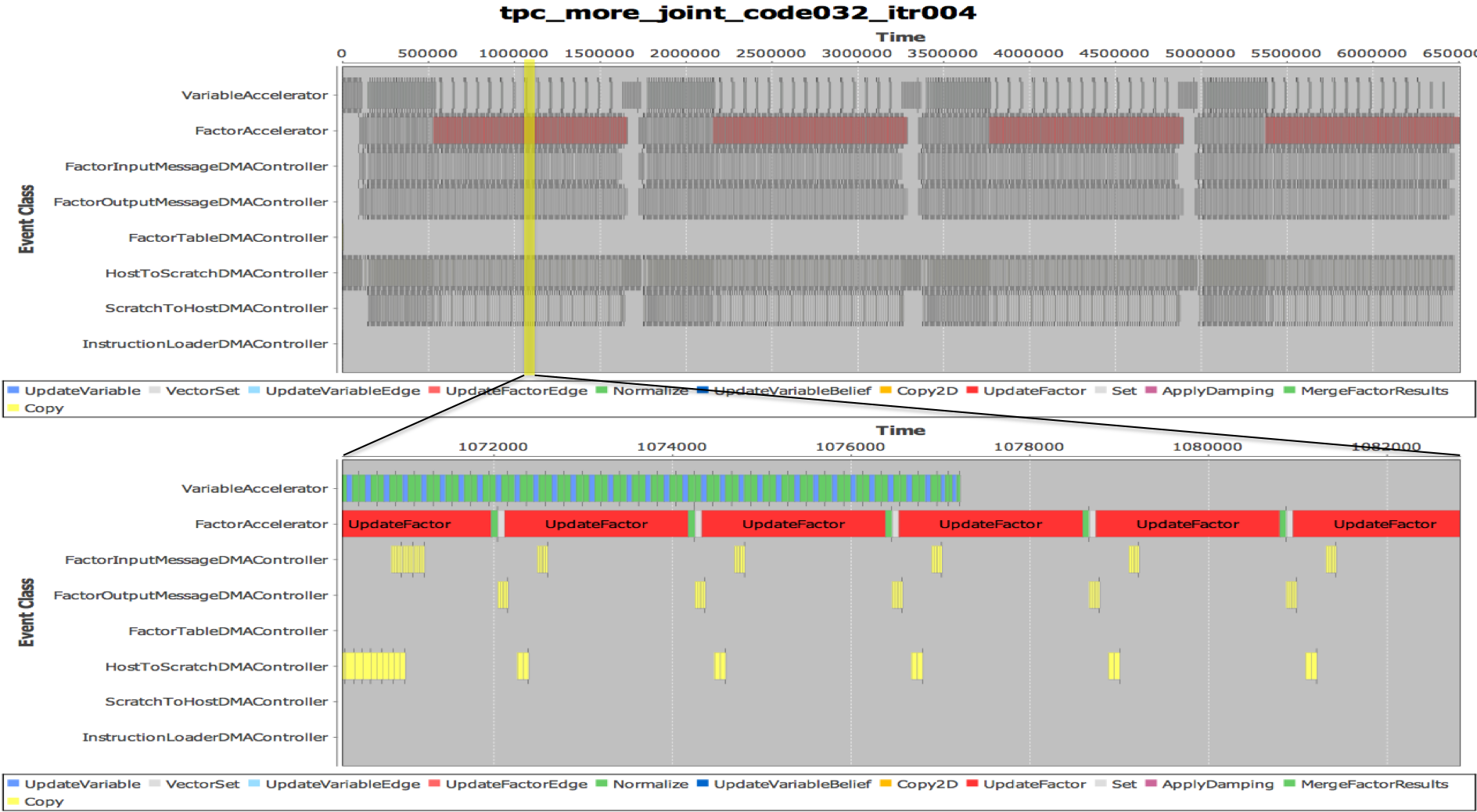}

\subsection{Dimple as a Compiler for GP5}
The GP5 hardware requires a compiler to translate a Dimple model to a set of GP5 hardware instructions.  The GP5 Compiler fits nicely within the Dimple architecture, acting as just another inference engine; as such, it is optimized for the GP5 hardware. The design problems associated with compiling down to the GP5 are unique and hard:  for normal compiler design, optimal allocation of variables to registers is an NP-complete map coloring problem;  GP5 faces an even harder version of this due to the requirement of allocating larger blocks of memory.  Additionally, CISC compilers provide challenges because different instructions might take different numbers of cycles.  In order to get maximum efficiency wins, timing of GP5 instructions varies over orders of magnitude.  
\section{Future Improvements}

Lyric has a long list of plans for improving Dimple.  Some of the immediate plans include speed improvements, more real variable support, integrating with languages like Stochastic MATLAB (an early prototype already exists), providing modelers in different languages, and tools to make it easier to provide inference engines in different languages.
\subsubsection*{References}

\small{
[1] S. Hershey, B. Vigoda, J. Bernstein, B. Bradley, T. Weber, A. Schweitzer, N. Stein Dimple, Lyric Labs Analog Devices, 2012. http://github.com/AnalogDevicesLyricLabs/dimple

[2] Noah D. Goodman, Vikash K. Mansinghka, Daniel M. Roy, Keith Bonawitz, and Joshua B. Tenenbaum.  
Church: a language for generative models. 
Proc. Uncertainty in Artificial Intelligence (UAI), 2008.

[3] T. Minka, J. Winn, J. Guiver, and D. Knowles
Infer.NET 2.4, Microsoft Research Cambridge, 2010.
http://research.microsoft.com/infernet

[4] Kevin Murphy, Bayes Net Toolbox, http://code.google.com/p/bnt

[5] Shaikh Faisal Zaheer, LDPC Code Simulation, http://www.mathworks.com/matlabcentral/fileexchange/8977-ldpc-code-simulation

[6]  S. Hershey, J. Bernstein, B. Bradley, T. Weber dimpleDemos, Lyric Labs Analog Devices, 2012. http://github.com/AnalogDevicesLyricLabs/dimpleDemos

[7] David Wingate, Andreas Stuhlmueller and Noah D. Goodman, Lightweight implementations of probabilistic programming languages via transformational compilation, Proc. of AISTATS, 2011.

\end{document}